\documentclass[preprint,aps,prd,showpacs,superscriptaddress,nofootinbib]{revtex4}
\usepackage{amsfonts}
\usepackage{mathrsfs}
\usepackage{graphicx}

\usepackage{amsmath}
\usepackage{amsthm}

\def\slash#1{#1 \hskip-0.45em /}

\begin{document}
\preprint{To be published in Chinese Physics C}

\title{\mbox{}\\[10pt]
The relativistic corrections to the fragmentation functions for a heavy quark to $B_c$ and $B_c^{*}$}


\author{Deshan Yang\footnote{E-mail: yangds@ucas.ac.cn}}
\affiliation{School of Physical Sciences, University of
Chinese Academy of Sciences, Beijing 100049, China\vspace{0.2cm}}
\affiliation{Institute of High Energy Physics, Chinese Academy of Sciences, Beijing 100049, China\vspace{0.2cm}}
\author{Wenjie Zhang\footnote{E-mail: zhangwenjie15@mails.ucas.ac.cn}}
\affiliation{School of Physical Sciences, University of
Chinese Academy of Sciences, Beijing 100049, China\vspace{0.2cm}}

\date{May 8, 2019}
\begin{abstract}
{In this paper, we compute the relativistic corrections to the fragmentation functions (FFs) of a heavy quark to $B_c$ and $B_c^*$ within the framework of non-relativistic QCD (NRQCD) factorization. The non-singlet and singlet DGLAP evolution are also presented.  }
\end{abstract}

\pacs{\it 12.38.-t, 12.38.Cy, 12.39.St, 14.40.Gx}

\maketitle

\section{Introduction\label{sect:intro}}   

One of main fields for precision examination of the perturbative Quantum Chromodynamics (QCD) is the study of the hadron production with the momentum $p$ ($p^2=m_H^2$) at the large transverse momentum $p_T$ where $p_T^2 >> m_H^2$. At the leading power of $1/p_T$, the fragmentation mechanism may be dominant so that the differential cross-section can be factorized as (for reviews, see \cite{Altarelli:1981ax,Collins:1989gx,Ellis:1991qj,Collins:2011zzd})
\begin{eqnarray}
E_p\frac{d\sigma_{A+B\to H+X}}{d^3p}\approx \sum\limits_f\int\frac{dz}{z^2}D_{f\to H}(z;\mu)E_c\frac{d\sigma_{A+B\to f(p_c)+X}}{d^3p}\left(p_c=\frac{1}{z}p\right)\,,
\end{eqnarray}
where parton-level differential cross-section $d\sigma_{A+B\to f(p_c)+X}$ contains the short-distance dynamics, and 
$D_{f\to H}(z;\mu)$ is the so-called single parton fragmentation function (FF) of the hadron $H$ from parton $f$ parametrizing the universal hadronization effects \cite{Collins:1981uk,Collins:1981uw}. Generally, the fragmentation functions are non-perturbative, but their renormalization scale-dependences are perturbatively calculable, and can be described by the famous Dokshitzer-Gribov-Lipatov-Altarelli-Parisi (DGLAP) equations which also govern the renormalization group (RG) running of the parton distribution functions (PDFs) 
\begin{eqnarray}
\label{eq:DGLAP}
\mu^2\frac{d}{d\mu^2}D_{f\to H}(z;\mu)&=&\sum\limits_{f^\prime}\int_z^1 \frac{dz^\prime}{z^\prime}P_{f\leftarrow f^\prime}(z/z^\prime,\alpha_s(\mu))D_{f^\prime\to H}(z^\prime;\mu)\,,
\end{eqnarray}
where $P_{f\leftarrow f^\prime}(z,\alpha_s(\mu))$ are the celebrated Altarelli-Parisi splitting functions \cite{DGLAP,LOAPkernel}.  

For the FFs of quarkonia, which are different from the FFs for the light mesons which rely completely on the dynamics in the non-perturbative regime of QCD, one believes that they can be further factorized into the products of the perturbatively calculable parts and non-perturbative behavior of the wave-functions of quarkonia at origin, due to the nature of quarkonium as a non-relativistic bound state of heavy quark and anti-quark. The standard theoretical tool to deal with the heavy quark bound state system is the NRQCD factorization \cite{Bodwin:1994jh}, in which all information of hadronization of quarkonium are encoded in the NRQCD matrix elements. In the literature \cite{Braaten:1996pv,Braaten:1993rw,Braaten:1994kd,Braaten:1996rp,Ma:1994zt,Sang:2009zz,Hao:2009fa}, various single-parton FFs quarkonia have been calculated. Recently, some of such fragmentations for quarkonia have been calculated to the higher order of $\alpha_s$ and $v$ analytically or numerically \cite{Ma:2013yla,Artoisenet:2014lpa,Zhang:2017xoj,Feng:2017cjk,Artoisenet:2018dbs,Feng:2018ulg,Zhang:2018mlo,Bodwin:2003wh,Bodwin:2002hg,Ma:2015lka} by implementing the state-of-art multi-loop calculation techniques.

The $B_{c}$ mesons are unique mesons which consist of two different flavor heavy quarks. Their production at high energy collision is a hot topic of perturbative QCD since 1990s \cite{Chang:1991bp,Chang:1992bb,Chang:1992pt,Braaten:1993jn,Cheung:1993pk,Chen:1993ii,Yuan:1994hn,Chang:2001pm,Chang:2014jca,Zheng:2015ixa,Zheng:2017xgj,Zheng:2019gnb,Qiao:2011yk,Qiao:2011zc,Jiang:2015jma,Jiang:2015pah}. Among these works, a lot of efforts have been paid on the FFs of $B_c$ mesons, and most of calculations were done within the framework of NRQCD factorization similar to corresponding calculations for quarkonia. It is worth noting that the first NLO calculation of the FF for a heavy-quark to $B_c^{(*)}$ meson, which is quite non-trivial, has been carried out in \cite{Zheng:2019gnb}.

In this paper, we compute the relativistic corrections to the FFs of a heavy-quark to $B_c$ and $B_c^*$ at the leading order of $\alpha_s$. These corrections are as equally important as the NLO QCD radiative corrections due to the NRQCD power counting rule. Similar results had appeared in the appendix of \cite{Sang:2009zz} as a by-product of calculation of FFs for quarkonia. The new things in this paper, that distinguish from results of  \cite{Sang:2009zz}, are: 1) we consider a different scheme to calculate the relativistic corrections to S-wave heavy-quark bound system, in which the mass of $B_c^{(*)}$ meson appears as an overall factor different from the scheme adopted in \cite{Sang:2009zz} that the binding energy is treated as a part of non-relativistic corrections; 2) we consider the contribution from the color-octet NRQCD operators; 
3) we calculate the FF for a heavy quark to the transversely polarized $B_c^*$; 4) we further investigate the DGLAP evolution effects.

The paper is organized in the following way. We first present the definition of the FFs for $b$ quark to $B_c^{(*)}$ following Collins and Soper, and  the desired NRQCD factorization formula in Sect.\ref{sect:FF&NRQCD}. In Sect.\ref{sect:compt}, we illustrate the matching procedure, present the explicit expressions for the short-distance distributions, and compare them with those in literature. In Sect.\ref{sect:DGLAP}, we present the non-singlet and singlet DGLAP evolution of the obtained FFs. Finally, we summarize in Sect.\ref{sect:summary}.

\section{Fragmentation Function and its NRQCD facotrization \label{sect:FF&NRQCD}}
   In this section, we briefly review the definition of the fragmentation function which was given by Collins and Soper\cite{Collins:1981uk,Collins:1981uw}, and present the NRQCD factorization formula for FFs of $b$-quark to $B_c^{(*)}$.

 We adopt the following notations for the decompositions of momenta: the 4-velocity of  $B_c$ is $v^{\mu}$ with $v^2=1$. We also use the same notation $v$ for the non-relativistic expansion parameter, which is typical size of the relative velocity of quark and anti-quark inside a $B_c$ meson. In the context, one should not confuse these two. We also introduce two light-like vectors $n^{\mu}$ and $\bar{n}^{\mu}$ such that $n^{2}=\bar{n}^{2}=0$ and $\bar{n}\cdot n=1$. Any 4-vector $a^{\mu}$ can be decomposed as $a^{\mu}=a^{+}\bar{n}^{\mu}+a^{-}{n}^{\mu}+a_{\perp}^{\mu}$($a^{+}\equiv n\cdot a$ and $a^{-}\equiv \bar{n} \cdot a$) with $n\cdot a_{\perp}=\bar{n}\cdot a_{\perp}=0$. Thus, $a\cdot b=a^{+}b^{-}+a^{-}b^{+}+a_{\perp}\cdot b_{\perp}$. For convenience, we set $v^{\mu}=v^{+}\bar{n}^{\mu}+v^{-}n^{\mu}$. Apparently $v^{+}v^{-}=1/2$.

  The $d$-dimensional fragmentation function for $b$ quark to $B_c^{(*)}$ meson is defined in the way given by Collins and Soper in \cite{Collins:1981uk,Collins:1981uw},  
     \begin{eqnarray}
\label{eq:FFbBc}
&&D_{b\to B_c^{(*)}}(z;\mu)\nonumber\\
&=&\frac{z^{d-3}}{4 N_c}\sum\limits_X\int\limits_{-\infty}^{\infty}\frac{dx^-}{2\pi} e^{-ip^+ x^-/z}{\rm Tr}\left[\slash n\langle 0\vert W(0) b(0) \vert B_c^{(*)}(p),X\rangle\langle B_c^{(*)}(p),X\vert  \bar  b(x^- n) W^\dag(x^-)\vert 0\rangle\right]\nonumber\\
&=&\frac{z^{d-3}}{4 N_c}\sum\limits_X \delta(p^+/z -p^+-p_X^+){\rm Tr}\left[\slash n\langle 0\vert W(0)b(0) \vert B_c^{(*)}(p),X\rangle\langle B_c^{(*)}(p),X\vert  \bar  b(0)W^\dag(0)\vert 0\rangle\right]\,,\nonumber\\
\end{eqnarray}
where $b(x)$ is the $b$ quark field in QCD, and the Wilson line along a light-like path is defined as
\begin{eqnarray}
	W(x^-)\,\equiv\,{\rm P}\exp\left(+i g_s\int^{\infty}_{x^-} ds
A^+(sn)\right)\,,
\end{eqnarray}
in which $g_s$ is the SU(3) gauge coupling and $ A_\mu (x)\equiv A^A_\mu(x) T^A$ ($T^A$s are the generators of SU(3) group in the fundamental representation). Since in this paper we never encounter the situation that we need the extra dimension to regulate the ultraviolet or infrared divergences, we always set $d=4$. 

Since $B_c^{(*)}$  can be regarded as a non-relativistic bound-state of $b$ and anti-$c$ quark, the contribution around the heavy-quark mass may be further factorized. Therefore, up to ${\cal O}(v^2)$, we may have the factorization formula within the NRQCD factorization framework, 
\begin{eqnarray}
	&&D_{b\to B_c(^{2s+1}S_J)}(z,\mu)\nonumber\\
	&=&2 m_{B_c}\Bigg\{d^{(0)}_{^{2s+1}S_J^{[1]}}(z,\mu)\frac{\langle  {\cal O}_1^{B_c}(^{2s+1}S_J)\rangle}{M^4}+d^{(0)}_{^{2s+1}S_J^{[8]}}(z,\mu)\frac{\langle  {\cal O}_8^{B_c}(^{2s+1}S_J)\rangle}{M^4}\nonumber\\ &&
	~~~~~~~~~+d^{(2)}_{^{2s+1}S_J^{[1]}}(z,\mu)\frac{\langle {\cal P}_1^{B_c}(^{2s+1}S_J)\rangle}{M^6}
	\Bigg\}
	+{\cal O}(v^4)\,,\label{eq:NRQCDfact}
\end{eqnarray}
where the mass scale $M\equiv m_b+m_c$ is introduced to balance the mass dimensions so that  the perturbatively calculable short-distance distributions $d(z,\mu)$ are dimensionless, and matrix-elements of the singlet and octet NRQCD operators are defined as
\begin{eqnarray}\label{eq:NRQCDops1}
	\langle {\cal O}_1^{B_c}(^1S_0)\rangle&=&\sum\limits_X\langle 0\vert\chi^\dag_c\psi_b \vert B_c+X\rangle\langle B_c+X\vert \psi_b^\dag\chi_c\vert 0 \rangle\,,\\
	\langle {\cal O}_8^{B_c}(^1S_0)\rangle &=&\sum\limits_X\langle 0\vert\chi^\dag_c T^A\psi_b \vert B_c+X\rangle\langle B_c+X\vert \psi_b^\dag T^A\chi_c\vert 0 \rangle\,,\\
\label{eq:NRQCDops2}
	\langle {\cal P}_1^{B_c}(^1S_0)\rangle&=&\frac{1}{2}\left[\sum\limits_X\langle 0\vert\chi^\dag_c\psi_b \vert B_c+X\rangle\langle B_c+X\vert \psi_b^\dag\left( -\frac{i}{2} \overleftrightarrow{\mathbf{D}}\right)^{2}\chi_c\vert 0 \rangle+{\rm h.c.}\right]\,,
\\
\label{eq:NRQCDops3}
	\langle {\cal O}_1^{B_c^*}(^3S_1)\rangle&=&\sum\limits_X\langle 0\vert\chi^\dag_c\sigma^i\psi_b \vert B_c^*+X\rangle\langle B_c^*+X\vert \psi_b^\dag\sigma^i\chi_c\vert 0 \rangle\,,\\
	\langle {\cal O}_8^{B_c^*}(^3S_1)\rangle &=&\sum\limits_X\langle 0\vert\chi^\dag_c T^A\sigma^i\psi_b \vert B_c+X\rangle\langle B_c+X\vert \psi_b^\dag T^A\sigma^i\chi_c\vert 0 \rangle\,,\\
\label{eq:NRQCDops4}
	\langle {\cal P}_1^{B_c^*}(^3S_1)\rangle&=&\frac{1}{2}\left[\sum\limits_X\langle 0\vert\chi^\dag_c\sigma^i\psi_b \vert B_c^*+X\rangle\langle B_c^*+X\vert \psi_b^\dag\sigma^i\left( -\frac{i}{2} \overleftrightarrow{\mathbf{D}}\right)^{2}\chi_c\vert 0 \rangle+{\rm h.c.}\right]\,,
\end{eqnarray}
where $\chi_c$ and $\psi_b$ are the two-component effective fields in NRQCD for $\bar c$ and $b$ quark, respectively, $\sigma^i$ is the $i$-th Pauli matrix, and $\mathbf{D}=\mathbf{\nabla}-ig_s\mathbf{A}$ is the covariant derivative. Note that all the $B_c^{(*)}$ states defined in (\ref{eq:NRQCDops1}-\ref{eq:NRQCDops4}) are non-relativistically normalized, meanwhile such states are relativistically normalized in the definition of FF in (\ref{eq:FFbBc}). The factor of $2 m_{B_c}$ in  the factorization formula (\ref{eq:FFbBc}) is just from the relativistic normalization condition for single particle states. 

By using the vacuum saturation approximation, at the leading order of $\alpha_s$, the matrix elements $\langle \mathcal{O}_1\rangle$ can be related to the radial wave function of S-wave $B_c$ meson at the origin $R_{S}(0)$  in the color-singlet model through
\begin{equation}
 \langle {\cal O}_1^{B_{c}}(^1S_0) \rangle \simeq \frac{N_{c}}{2\pi} \vert R_{S}(0) \vert^{2}\,,~~~~\langle {\cal O}_1^{B_{c}}(^3S_1) \rangle \simeq (d-1)\frac{N_{c}}{2\pi} \vert R_{S}(0) \vert^{2}\,.
\end{equation}
However, there is no similar simple relation for the matrix-elements $\langle \mathcal{O}_8\rangle$.

\section{Computation of the Fragmentation Functions\label{sect:compt}}

\subsection{Matching procedure}
In the practical computation of the short-distance distributions $d(z,\mu)$, we replace the $B_c^{(*)}$ with a color-singlet quark pair $[b\bar c(^{2s+1}S_J^{[1]})]$, 
\begin{eqnarray}
	\vert [b\bar c(^{2s+1}S_J^{[1]})](p)\rangle =\sum\limits_{a,b=1}^{N_c}\frac{\delta_{ab}}{\sqrt{N_c}}\vert b^a(p_b),\bar c^b(p_{c})\rangle \,,
	\end{eqnarray}
	or a color-octet quark pair $[b\bar c(^{2s+1}S_J^{[8]})]$
\begin{eqnarray}
	\vert [b\bar c(^{2s+1}S_J^{[8]})](p)\rangle =\sum\limits_{a,b=1}^{N_c}\frac{T^A_{ab}}{\sqrt{T_F}}\vert b^a(p_b),\bar c^b(p_{c})\rangle \,,
	\end{eqnarray}	
	where $a,b$ are color indices, $T_F=1/2$ is the Dynkin index of SU(3) group in the fundamental representation. And we set the momenta as
\begin{eqnarray}
	p^\mu &=&M_H v^\mu =p_b^\mu+p_c^\mu\,,~~~p_b^2=m_b^2\,,~~p_c^2=m_c^2\,,\nonumber\\
	p_b&=&m_b v^\mu+k^\mu\,,~~p_c=m_c v^\mu+\tilde{k}^\mu\,,
\end{eqnarray}
where $k^\mu$ and $\tilde{k}^\mu$ are the residual momenta of $b$-quark and $\bar c$-quark, respectively. 

At leading order of $\alpha_s$, we have
\begin{eqnarray}
\label{eq:FFbc}
D_{b\to [b\bar c]}(z;\mu)=\frac{z}{4 N_c} \sum_{\text{colors\&spins}}\int\frac{dq^+ d^2 q_\perp}{(2\pi)^3 2 q^+} \theta(q^+)\delta(p^+/z-p^+-q^+) \left\vert {\cal M}\right\vert^2 
\,,\end{eqnarray}
with
\begin{eqnarray}
\left\vert {\cal M}\right\vert^2={\rm Tr}\left[\slash n\langle 0\vert W(0)b(0) \vert [b\bar c](p),c(q)\rangle\langle [b\bar c,c(q)](p),c(q)\vert   \bar  b(0)W^\dag(0)\vert 0\rangle\right]\,,
\end{eqnarray}
where $q^2=m_c^2$. Schematically, it has the similar NRQCD factorization formula as in (\ref{eq:NRQCDfact}) by just replacing the $B_c^{(*)}$ state by corresponding $b\bar c$ pair state.

 In the light-cone gauge, 
\begin{eqnarray}
&&\langle b^a(p_b),\bar c^b(p_{c}),c^c(q)\vert   \left[\bar  b(0)W^\dag(0)\right]_d\vert 0\rangle	\nonumber\\
&=&-g_s^2 T^A_{ad} T^A_{cb}\left\{\frac{\bar u_c(q)\gamma^\mu v_c(p_c)\bar u_b(p_b)\gamma_\mu (\slash p+\slash q+m_b)}{[(p+q)^2-m_b^2][(p_c+q)^2]}-\frac{\bar u_c(q)\slash n v_c(p_c)\bar u_b(p_b)}{[(p_c+q)^2][(p_c+q)^+]}\right\}\,,\label{eq:ME}
\end{eqnarray}
where we have used various equations of motion to simplify the spin structures. We also checked the expression of this matrix-element in the covariant gauge, which is the same as the above expression in the light-cone gauge as it should be.

Since the FF is defined in a Lorenz invariant way, we can do the calculation in any reference frame. For convenience, we choose the rest frame of the $b\bar c$ pair, in which we have $v^\mu=(1,\mathbf{0})$ and 
\begin{eqnarray*}
	k^\mu=\left(\frac{\mathbf{k}^2}{2 m_b},\mathbf{k}\right)\,,~~~\tilde k^\mu=\left(\frac{\mathbf{k}^2}{2 m_c},-\mathbf{k}\right)\,.
\end{eqnarray*}

Generally, one applies the replacements
	\begin{eqnarray}
 v(p_c)\bar u(p_b) &\to &\Pi_0(p_b,p_c)=i\frac{\left(
  \slash p_c-m_c\right)\gamma_5
  (\slash p+E_b+E_c)\left(\slash p_b+m_b\right)}{2\sqrt{2 E_b
  E_c}\sqrt{(E_c+m_c)(E_b+m_b)}{2(E_b+E_c)}}\,,\\
    v(p_c)\bar u(p_b)  &\to & \Pi_1(p_b,p_c)=-\frac{\left(
\slash p_c-m_c\right) \slash\varepsilon^*(
 \slash p+E_b+E_c)\left( \slash p_b+m_b\right)}{2\sqrt{2 E_b
  E_c}\sqrt{(E_c+m_c)(E_b+m_b)}{2(E_b+E_c)}}\,,\label{proj2}
\end{eqnarray}
to project out the spin-singlet and spin-triplet parts, respectively. Here $\varepsilon^\mu$ denotes the polarization vector for the spin-triplet quark pair, and 
the energies $E_b=\sqrt{m_b^2+\mathbf{k}^2}$ and $E_c=\sqrt{m_c^2+\mathbf{k}^2}$. After expanding the resulting matrix-element (\ref{eq:ME}) to the second order in the momentum ${\bf k}$, one can extract the S-wave contributions by neglecting the first order terms in ${\bf k}$ (which contribute to $P$-wave state only) and making the replacement $k^i k^j\to {\bf k}^2 \delta^{ij}/3$. 
 This leads to the corresponding matrix-elements of NRQCD operators
at leading order of $\alpha_{s}$, 
\begin{eqnarray}
 \langle {\cal O}^{b\bar{c}}_1 (^{1}{}S_{0}) \rangle &\equiv & \sum\limits_X\langle 0\vert\chi^\dag_c\psi_b \vert [b\bar c](^1S_0^{[1]})+X\rangle\langle [b\bar c](^1S_0^{[1]})+X\vert \psi_b^\dag\chi_c\vert 0 \rangle = 2N_{c},  \\
  \langle {\cal O}^{b\bar{c}}_8 (^{1}{}S_{0}) \rangle &\equiv & \sum\limits_X\langle 0\vert\chi^\dag_c\psi_b \vert [b\bar c](^1S_0^{[8]})+X\rangle\langle [b\bar c](^1S_0^{[8]})+X\vert \psi_b^\dag\chi_c\vert 0 \rangle =2 N_{c} C_F,  \\
 \langle {\cal O}^{b\bar{c}}_1 (^{3}{}S_{1}) \rangle &\equiv & \sum\limits_X\langle 0\vert\chi^\dag_c\sigma^i\psi_b \vert [b\bar c](^3S_1^{[1]})+X\rangle\langle [b\bar c](^3S_1^{[1]})+X\vert \psi_b^\dag\sigma^i\chi_c\vert 0 \rangle =2(d-1)N_{c} , \\
  \langle {\cal O}^{b\bar{c}}_8 (^{3}{}S_{1}) \rangle &\equiv & \sum\limits_X\langle 0\vert\chi^\dag_c\sigma^i\psi_b \vert [b\bar c](^3S_1^{[8]})+X\rangle\langle [b\bar c](^3S_1^{[8]})+X\vert \psi_b^\dag\sigma^i\chi_c\vert 0 \rangle =2(d-1)N_{c}C_F , \nonumber\\\\
 \langle {\cal P}^{b\bar{c}}_1(n) \rangle &=&  \mathbf{k}^{2} \langle{\cal O}^{b\bar{c}}_1(n) \rangle,~~n=^1S_0^{[1]}\,,^3S_1^{[1]}\,.
\end{eqnarray}

  It is worth noting that we calculate the FF for $B_{c}^{*}$ by using the replacement rules for the summation of polarization vectors
  \begin{equation}
  \sum\limits_{\lambda=0,\pm 1}\varepsilon^{*\mu}(p,\lambda)\varepsilon^{\nu}(p,\lambda) \to -g^{\mu\nu} + \frac{p^{\mu}p^{\nu}}{p^2} = -g^{\mu\nu} + v^{\mu}v^{\nu},
  \end{equation}
 for the unpolarized case, and
  \begin{equation}
   \sum\limits_{\lambda=\pm 1}\varepsilon^{*\mu}(p,\lambda)\varepsilon^{\nu}(p,\lambda) \to -g^{\mu\nu} + \frac{p^{\mu}n^{\nu} + p^{\nu}n^{\mu}}{p^{+}} - \frac{p^{2} n^{\mu} n^{\nu}}{(p^{+})^2} = -g^{\mu\nu} + \frac{v^{\mu}n^{\nu} + v^{\nu}n^{\mu}}{ v^+} - \frac{n^{\mu}n^{\nu}}{(v^+)^{2}}\,,
  \end{equation}
for the transversely polarized case. We will invoke a superscript T for FF and corresponding short-distance distributions of transversely polarized $B_c^*$.

Through the matching equations,
\begin{eqnarray}
	 D_{b\to b\bar{c}(^{1}{}S_{0}^{[1]})} & =& \frac{2 N_{c}}{M^4}\left( d^{(0)}_{^{1}{}S_{0}^{[1]}} + \frac{\mathbf{k}^{2}}{M^2} d^{(2)}_{^{1}{}S_{0}^{[1]}}\right),  \\
  D_{b\to b\bar{c}(^{3}{}S_{1}^{[1]})} & =& \frac{2 (d-1)N_{c}}{M^4}\left( d^{(0)}_{ ^{3}{}S_{1}^{[1]}} + \frac{\mathbf{k}^{2}}{M^2} d^{(2)}_{ ^{3}{}S_{1}^{[1]}}\right),\\
   D_{b\to b\bar{c}(^{1}{}S_{0}^{[8]})} & =& \frac{2 N_{c}C_F}{M^4} d^{(0)}_{ ^{1}{}S_{0}^{[8]}} ,  \\
  D_{b\to b\bar{c}(^{3}{}S_{1}^{[8]})} & =& \frac{2 (d-1)N_{c}C_F}{M^4} d^{(0)}_{ ^{3}{}S_{1}^{[8]}} ,
\end{eqnarray}
we can extract the short-distance distributions straightforwardly.

\subsection{Results of short distance distributions}

At the lowest order of $v$,
\begin{eqnarray}
   d^{(0)}_{ ^{1}{}S_{0}^{[1]}}(r, z) \nonumber
&=& \frac{C_F^2 \alpha_{s}^{2}}{24 N_c^2}\frac{z(1-z)^{2}}{r^2(1-\bar r z)^{6}} \Bigg[6-18(1-2r)z+(21-74r+68r^{2})z^{2}  \\
&& -2\bar r(6-19r+18r^{2})z^{3}+3\bar r^{2}(1-2r+2r^{2})z^{4}\Bigg],\\
 d^{(0)}_{ ^{3}{}S_{1}^{[1]}}(r, z)  \nonumber
&=& \frac{C_F^2 \alpha_{s}^{2}}{24 N_c^2}\frac{z(1-z)^{2}}{r^2 (1-\bar r z)^{6}} \Bigg[2-2(3-2r)z +3(3-2r+4r^2)z^2  \\
&&  -2\bar r(4-r+2r^2)z^3 +\bar r^{2}(3-2r+2r^2)z^4\Bigg]\,,\\
 d^{(0)}_{ ^{1}{}S_{0}^{[8]}}(r, z)&=&  \frac{1}{2N_cC_F^2} d^{(0)}_{b \rightarrow ^{1}{}S_{0}^{[1]}}(r, z) \,,\\
d^{(0)}_{^{3}{}S_{1}^{[8]}}(r, z)&=& \frac{1}{2N_c C_F^2}d^{(0)}_{b \rightarrow ^{3}{}S_{1}^{[1]}}(r, z),
\end{eqnarray}
in which we denote $r\equiv m_c/(m_b+m_c)$ and $\bar r\equiv 1-r$ for shortening the expressions. The color factor $C_F=(N_c^2-1)/(2 N_c)$ with $N_c=3$. Here we fix $\alpha_s$ at $\mu=m_b+m_c$. 
$d^{(0)}_{^{1}{}S_{0}^{[1]}}(r, z)$ and $d^{(0)}_{ ^{3}{}S_{1}^{[1]}}(r, z)$ agree with the results obtained by Braaten et al in \cite{Braaten:1993jn} by expressing $\langle O_1\rangle$ in term of the Schr\"odinger wave function at origin.

At the order of $v^2$, we get
\begin{eqnarray}
 d^{(2)}_{^{1}{}S_{0}^{[1]}}(r, z) &=& -\frac{ \alpha_{s}^{2}C_F^2}{288N_c^2} \frac{z(1-z)^{2} }{r^{4}\bar r^{2} \left( 1-\bar rz\right)^{8}}\Bigg[66-48 r+ \left(96 r^3-504 r^2+792 r-330\right) z \nonumber \\
&& +\left(904 r^4-2624 r^3+3810 r^2-2738 r+693\right) z^2 \nonumber \\
&&  -8 \bar r \left(316 r^4-576
   r^3+670 r^2-426 r+99\right) z^3  \nonumber \\
&& +2 \bar r^2 \left(1064 r^4-1744
   r^3+1919 r^2-1158
   r+264\right) z^4 \nonumber\\
&&  -2 \bar r^3 \left(336 r^4-524
   r^3+664 r^2-415 r+99\right) z^5 \nonumber\\
&&  +3 \bar r^4 \left(24 r^4-32
   r^3+58 r^2-42 r+11\right) z^6 \Bigg],
\end{eqnarray}
and
\begin{eqnarray}
   d^{(2)}_{ ^{3}{}S_{1}^{[1]}}(r, z)
&= &- \frac{\alpha_{s}^{2}C_F^2}{864 N_c^2} \frac{ z (1-z)^{2} }{r^4\bar r^{2} (1-\bar r z)^{8}}\Bigg[66-96 r+48 r^2 +\left(384 r^3-840 r^2+768 r-330\right) z  \nonumber\\
  && + \left(472 r^4-2432 r^3+3466 r^2-2130 r+759\right) z^2  \nonumber\\
  && - 8 \bar r \left(196 r^4-280
   r^3+492 r^2-291
   r+132\right) z^3 \nonumber\\
  && + 2 \bar r^2 \left(560 r^4-488
   r^3+1383 r^2-882
   r+462\right) z^4 \nonumber\\
  && -2 \bar r^3 \left(192 r^4-164
   r^3+692 r^2-471
   r+231\right) z^5 \nonumber\\
  && + 3 \bar r^4 \left(40 r^4-64
   r^3+130 r^2-82 r+33\right) z^6 \Bigg].
\end{eqnarray}

For transversely polarized fragmentation function for $B_c^{*}$, we have the short-distance distributions as
\begin{eqnarray}
   d^{{\rm T}(0)}_{^{3}{}S_{1}^{[1]}}(r, z) 
&= & \frac{ \alpha_{s}^{2}C_F^2}{36 N_c^2}\frac{z(1-z)^2 }{r^2 [1-\bar r z]^6}\Bigg[  2 +(4 r-6) z +\left(10 r^2-4
   r+9\right) z^2 -2 \bar r \left(r+4\right) z^3+3 \bar r^2 z^4\Bigg],\nonumber\\\\
    d^{{\rm T}(0)}_{ ^{3}{}S_{1}^{[8]}}(r, z)&=&\frac{1}{2N_cC_F^2} d^{{\rm T}(0)}_{ b \rightarrow ^{3}{}S_{1}^{[1]}}(r, z)\,,
   \end{eqnarray}
and   
   \begin{eqnarray}
   d^{{\rm T}(2)}_{^{3}{}S_{1}^{[1]}}(r, z) 
&=& -\frac{\alpha_{s}^{2}C_F^2}{864N_c^2 }\frac{ z(1-z)^2  }{r^4\bar r^2 (1-\bar r z)^8} \Bigg[ 22 -32 r+ 16 r^2+2 \left(64 r^3-140 r^2+128 r-55\right) z  \nonumber\\
  &&  +\left(176 r^4-832 r^3+1140 r^2-692 r+253\right) z^2  \nonumber\\
  &&  -8\bar r\left(64 r^4-90 r^3+153
   r^2-88 r+44\right) z^3  \nonumber\\
  &&  +2 \bar r^2 \left(128 r^4-72 r^3+375 r^2-240 r+154\right) z^4 \nonumber\\
  &&  -2 \bar r^3 \left(44 r^3+160 r^2-121r+77\right) z^5 + \bar r^4 \left(88 r^2-64 r+33\right) z^6 \Bigg].
\end{eqnarray}


In \cite{Braaten:1993jn}, the authors attribute $\int_0^z D_{b\to B_c^{(*)}}(z)$ to be a kind of fragmentation probability. It is quite interesting to investigate that the ratios among such fragmentation probabilities of $b$ to $B_c$, unpolarized $B_c^{*}$ and transversely polarized $B_c^{*}$, such as
\begin{eqnarray*}
	R_{13}(r)&=&\frac{\int_0^1 dz \,D_{b\to B_c}(r,z)}{\int_0^1 dz \,D_{b\to B_c^*}(r,z)}\approx\frac{1}{3}\frac{\int_0^1 dz \,d^{(0)}_{^1S_0}(r,z)}{\int_0^1 dz \,d^{(0)}_{^3S_1}(r,z)}=\frac{1}{3}-\left(\frac{35}{36}-\frac{5 \ln r}{6}\right) r 
	+{\cal O}(r^2)\,,\nonumber\\
	R_{{\rm T}3}(r)&=&\frac{\int_0^1 dz \,D^{\rm T}_{b\to B_c}(r,z)}{\int_0^1 dz \,D_{b\to B_c^*}(r,z)}\approx\frac{\int_0^1 dz \,d^{{\rm T}(0)}_{^3S_1}(r,z)}{\int_0^1 dz \,d^{(0)}_{^3S_1}(r,z)}=\frac{2}{3}+\frac{5}{18} r
	+{\cal O}(r^2)\,.
	\end{eqnarray*}
 One can see that such ratios in the limit $r\to 0$ agree with the rough estimation based on arguments of counting the degrees of freedom on polarizations. Even when ${\cal O}(v^2)$ corrections are taken into account, the above limit of ratios when $r$ goes to zero still hold. 

The fragmentation functions $D_{\bar c\to B_c^{(*)}}$ can be easily obtained by replacing $r$ to $1-r$ in the above results. Namely, one has
\[
D_{\bar c\to B_c^{(*)}}(r,z)=D_{b\to B_c^{(*)}}(\bar r,z)\,,\]
which is just a consequence of a kind of charge charge-conjugation \cite{Braaten:1993jn}.

\subsection{Comparisons and discussions}
 
 As mentioned in Sect.\ref{sect:intro}, in \cite{Sang:2009zz}, the relativistic corrections to the FF have been calculated. Neglecting the color-octet contribution, and taking into account the Gremm-Kapustin relation \cite{Gremm:1997dq}, that is for $B_c^{(*)}$,
\begin{align}
 M_{B_{c}^{(*)}} = m_b+m_c + \frac{\langle \mathbf{k}^{2} \rangle_{B_{c}^{(*)}}}{2m_b}+\frac{\langle \mathbf{k}^{2} \rangle_{B_{c}^{(*)}}}{2m_c} + {\cal O}(v^4),
\end{align}
where 
\begin{eqnarray}
\left\langle \mathbf{k}^{2} \right\rangle_{B_{c}^{(*)}}\equiv \frac{\left\langle {\cal P}_1^{B_c^{(*)}}
\right\rangle}{\left\langle {\cal O}_1^{B_c^{(*)}}\right\rangle}	\,,
\end{eqnarray}
we can derive the same results as in \cite{Sang:2009zz}. 

We note that in many phenomenological applications of NRQCD factorization, the matrix-element $\langle {\cal P}^H\rangle$ is treated as a free parameter which is usually not bounded by the Gremm-Kapustin relation. In this sense, it is inconsistent with attributing the binding energy as part of relativistic corrections, which is equivalent to apply the Gremm-Kapustin relation, in the matching procedure. Our scheme to calculate the relativistic corrections by taking out the meson mass as an overall factor is more relevant and consistent in phenomenological applications. 

\section{Evolution of Fragmentation Function\label{sect:DGLAP}}
In this section, we solve the DGLAP evolution equations numerically by using the 4-th order classical Runge-Kutta method. We consider both the non-singlet and singlet evolution. We neglect the contributions from the color-octet operators.

The singlet DGLAP equations read as 
 \begin{equation}
  \mu \frac{\partial}{\partial \mu}D_{f\rightarrow B_{c}}(z;\mu)= \frac{\alpha_{s}(\mu^{2})}{\pi}\sum_{f' \in ( b, \bar{c},g)}\int_{z}^{1}\frac{dz'}{z'}P_{f'\leftarrow f}(z/z',\alpha_{s}(\mu))D_{f'\rightarrow B_{c}}(z';\mu)\,,~~~~f=b,\bar c, g,
  \end{equation}
 where the Altarali-Parisi kernels $ P_{f'\leftarrow f}(z/z',\alpha_{s}(\mu)) $ are
 \begin{align}
 P_{q\leftarrow q}(z) &= C_{F}\left[\frac{1+z^2}{(1-z)_{+}}+\frac{3}{2}\delta (1-z)\right]+{\cal O}(\alpha_{s}^{2}), \\
 P_{g\leftarrow q}(z) &= C_{F}\left[\frac{1+(1-z)^{2}}{z}\right]+{\cal O}(\alpha_{s}^{2}), \\
 P_{q\leftarrow g}(z) &= T_F[z^{2}+(1-z)^{2}]+{\cal O}(\alpha_{s}^{2}), \\
 P_{g\leftarrow g}(z) &= \left\{2C_{A}\left[\frac{z}{(1-z)_{+}}+(1-z)(z+\frac{1}{z})\right]+\frac{\beta_{0}}{2}\delta(1-z)\right\}+{\cal O}(\alpha_{s}^{2}).
 \end{align}
where $C_A=N_c$ and $\beta_{0}=11C_{A}/3-4 T_F n_{f}/3$. The non-singlet DGLAP equation can be obtained from the singlet case by neglecting the off-diagonal Altarali-Parisi kernels. 

Here the sliding scale $\mu$ is rather as factorization scale than renormalization scale since eventually the such scale dependence of fragmentation functions will be canceled by the corresponding factorization scale dependence in the parton-level cross-sections or partial decay widths after convolution. Of course, the fragmentation function of heavy-quark to $B_c$ will be also renormalization scale dependent. However, such dependence should be canceled by the $\beta$-function dependent terms (mainly from virtual vertex corrections and self-energy corrections) in the higher order radiative corrections to these fragmentation functions, which is not the main concern of this work.  

\begin{figure}[t]
	\centerline{\includegraphics[width=4.5in]{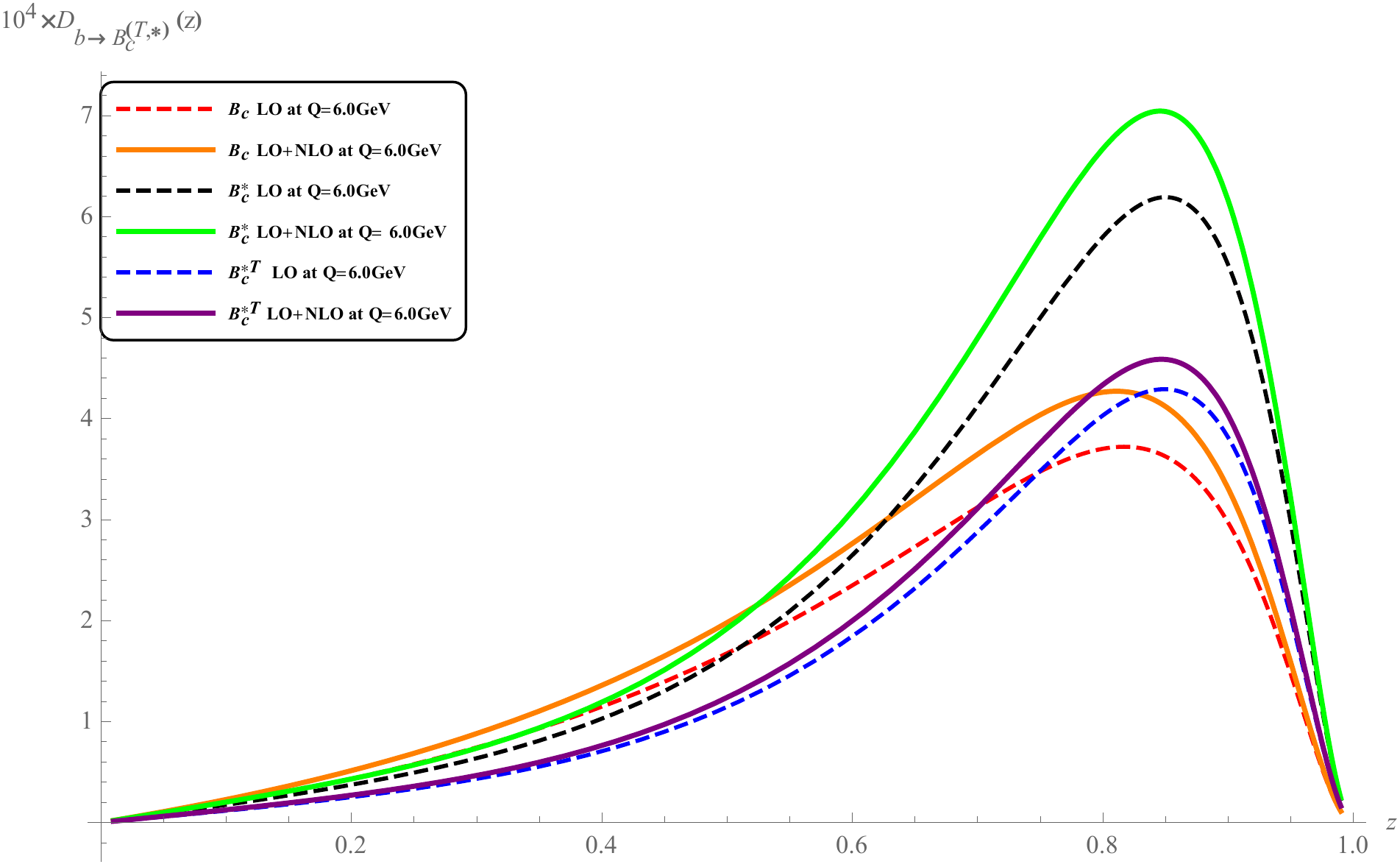}}
	\caption{The FFs $D_{b\to B_c}$, $D_{b\to B_c^*}$ and $D_{b\to B_c^*}^{\rm T}$ at $\mu=6$ GeV without/with relativistic corrections.\label{fig:FFsat6GeV}}
\end{figure}

\begin{figure}[t]
	\centerline{\includegraphics[width=4.5in]{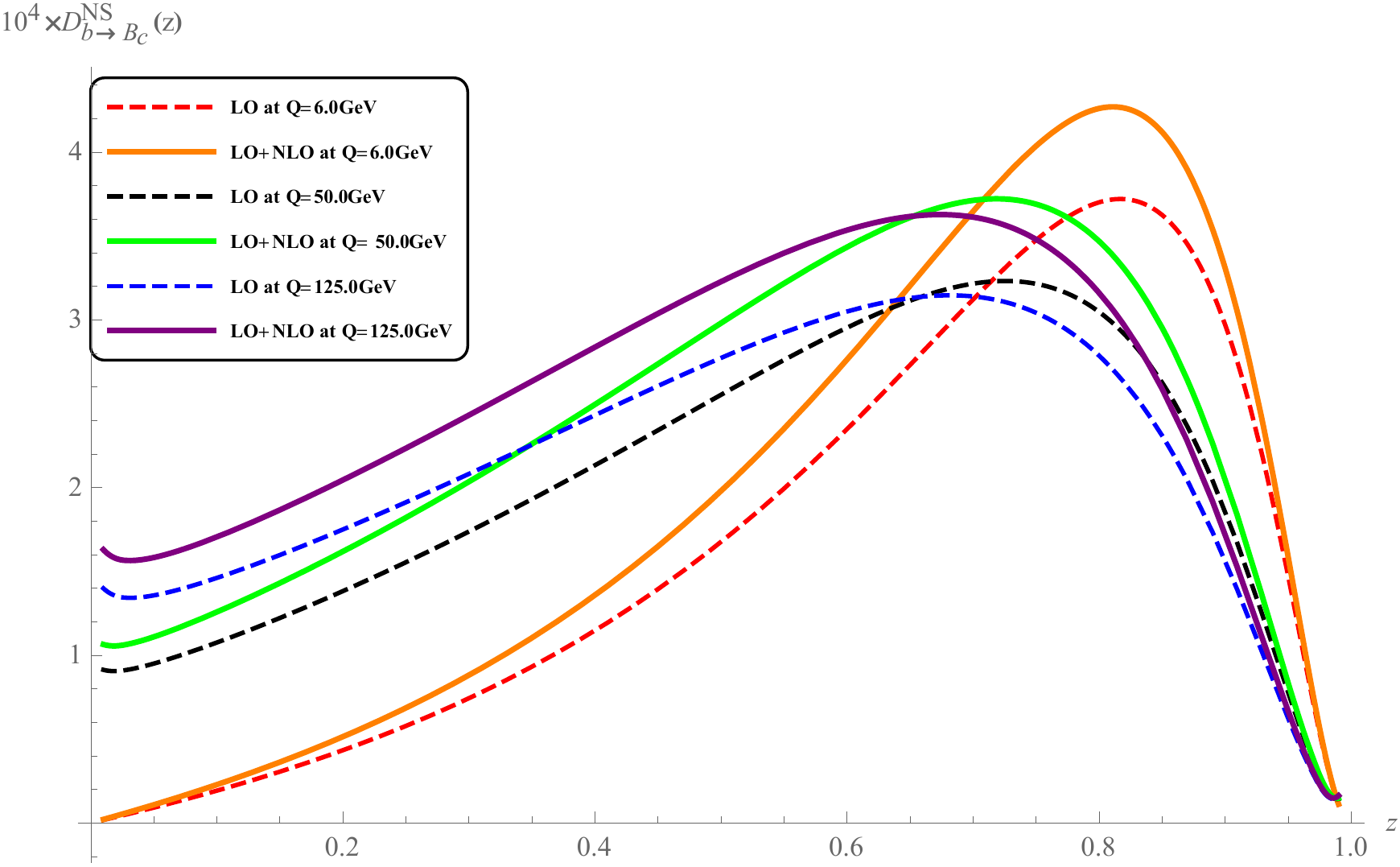}}
	\caption{The FFs $D_{b\to B_c}$ at $\mu=6.0$ GeV, $\mu=50.0$ GeV and $\mu=125.0$ GeV without/with relativistic corrections under non-singlet DGLAP evolution.\label{fig:FFBcNSDGLAP}}
\end{figure}

\begin{figure}[t]
	\centerline{\includegraphics[width=4.5in]{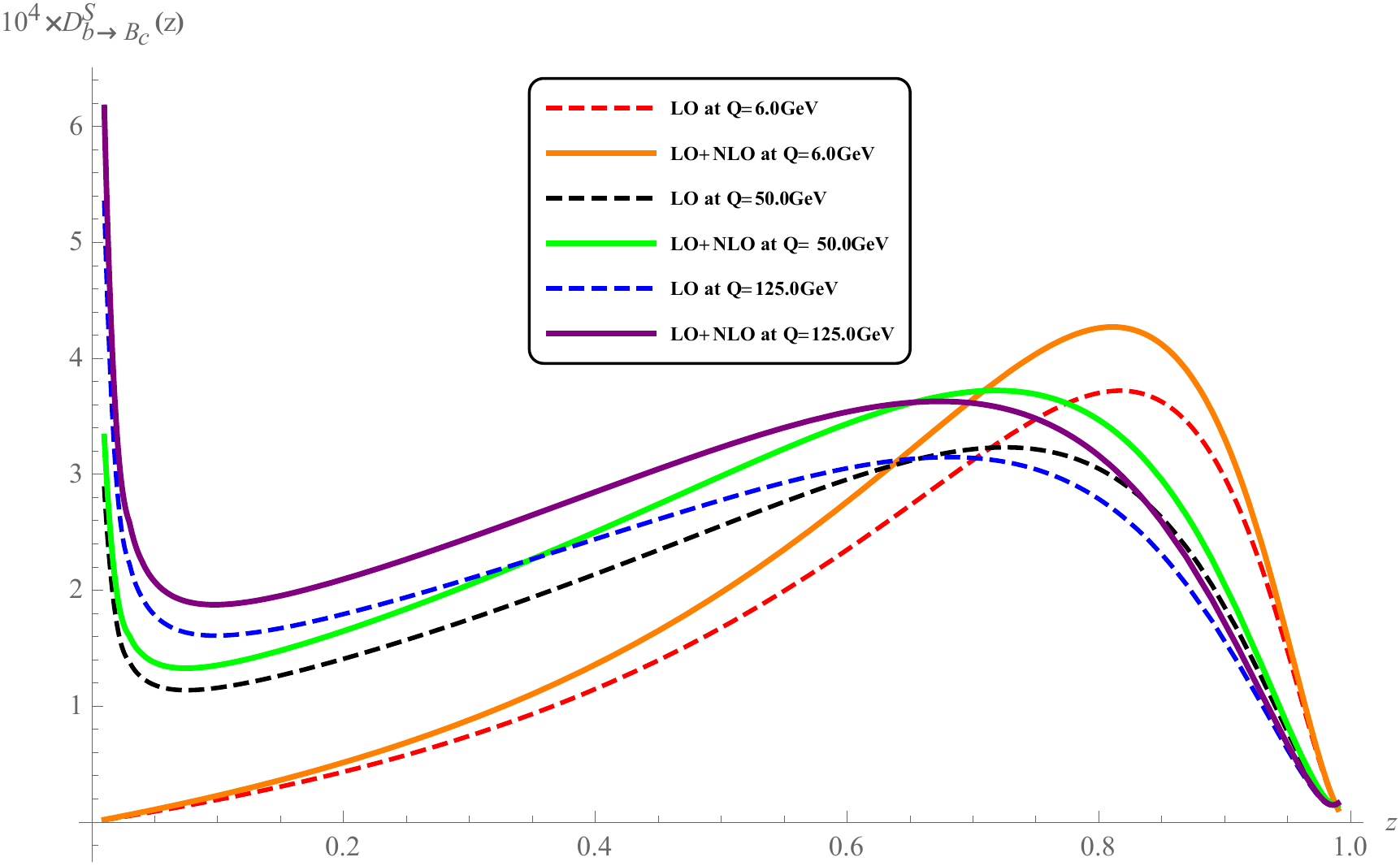}}
	\caption{The FF $D_{b\to B_c}$ at $\mu=6.0$ GeV, $\mu=50.0$ GeV and $\mu=125.0$ GeV without/with relativistic corrections under singlet DGLAP evolution.\label{fig:FFbtoBcSDGLAP}}
\end{figure}

\begin{figure}[t]
	\centerline{\includegraphics[width=4.5in]{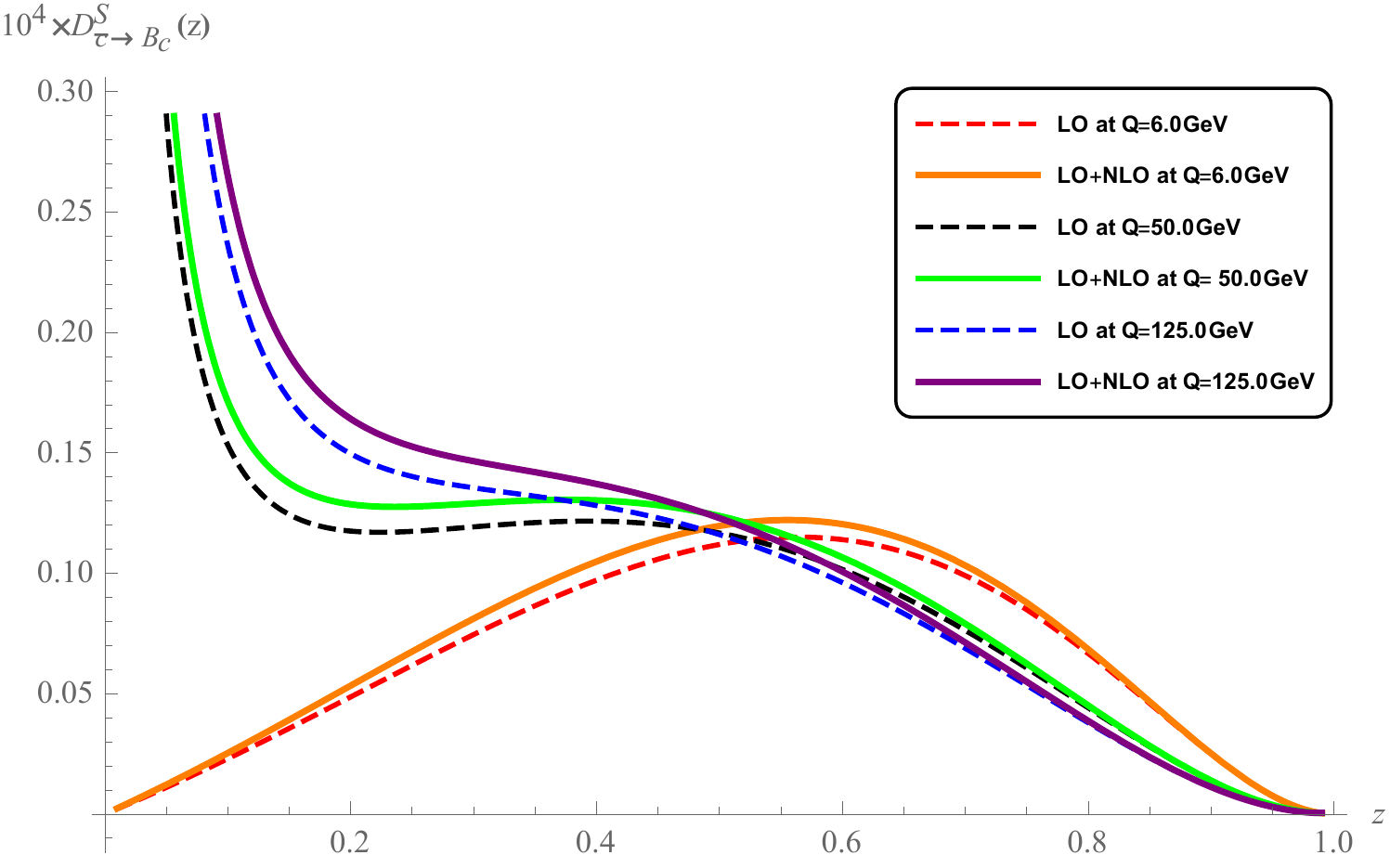}}
	\caption{The FF $D_{\bar c\to B_c}$ at $\mu=6.0$ GeV, $\mu=50.0$ GeV and $\mu=125.0$ GeV without/with relativistic corrections under singlet DGLAP evolution.\label{fig:FFcbartoBcSDGLAP}}
\end{figure}

\begin{figure}[t]
	\centerline{\includegraphics[width=4.5in]{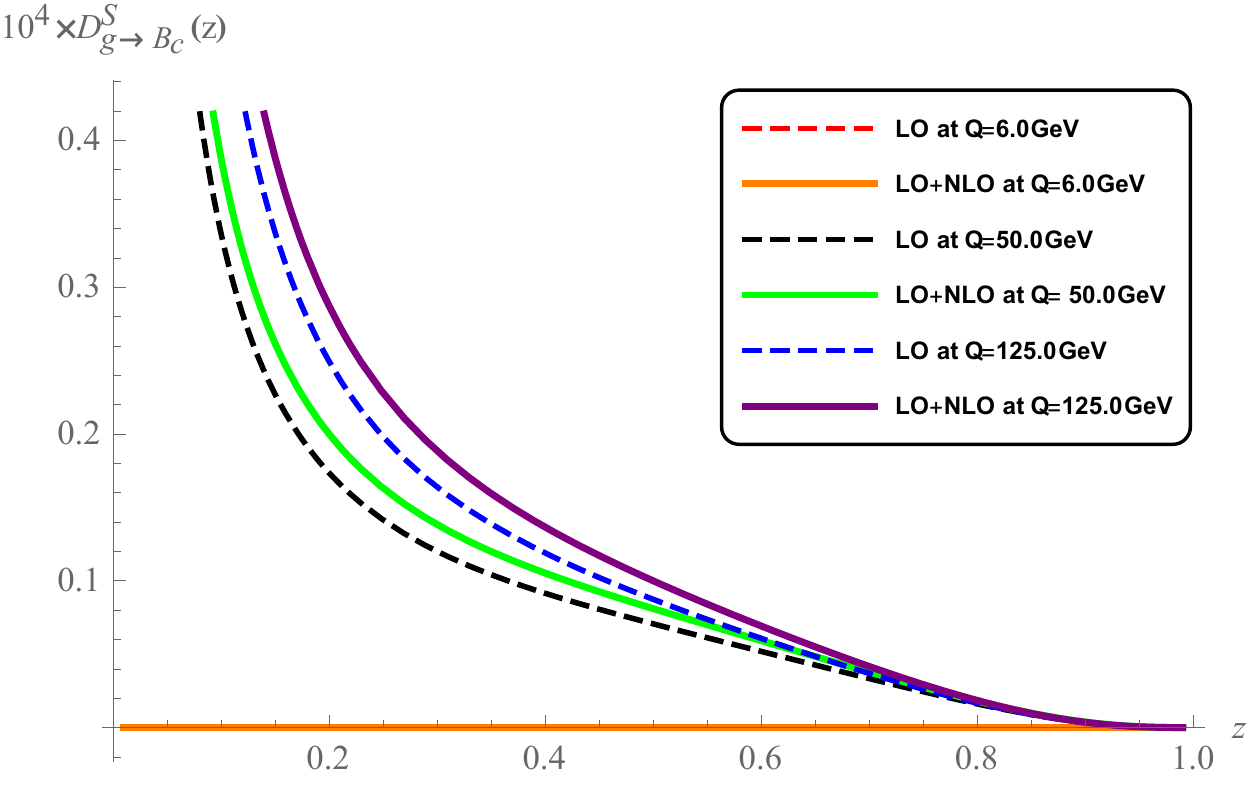}}
	\caption{The FF $D_{g\to B_c}$ at $\mu=6.0$ GeV, $\mu=50.0$ GeV and $\mu=125.0$ GeV without/with relativistic corrections under singlet DGLAP evolution.\label{fig:FFgtoBcSDGLAP}}
\end{figure}

We take the strong coupling $\alpha_s(m_Z)=0.118$ with the leading logarithm renormalization running with $n_f=5$, the heavy quark pole masses $m_b=4.78$ GeV and $m_c=1.67$ GeV, $m_{B_c}=6.275$ GeV \cite{Tanabashi:2018oca}. And we also take the Schr\"odinger wave function at origin for $B_c^{(*)}$ from the potential model in \cite{Eichten:1994gt}, which is $\vert R_S(0)\vert^2=1.642\,\text{GeV}^3$. This implies that
\begin{eqnarray*}
	\langle \mathcal{O}_1^{B_c}(^1S_0)\rangle \simeq 0.784\,\text{GeV}^3\,,~~~~\langle \mathcal{O}_1^{B_c^*}(^3S_1)\rangle \simeq 2.346\,\text{GeV}^3\,.
\end{eqnarray*}
We use the Gremm-Kapustin relation and heavy-quark spin symmetry to get the values for
\begin{eqnarray}
	\langle \mathcal{P}_1^{B_c}(^1S_0)\rangle &\simeq & \frac{2m_b m_c(M_{B_c}-m_b-m_c)}{m_b+m_c}\langle \mathcal{O}_1^{B_c}(^1S_0)\rangle=-0.330\,\text{GeV}^5\,,\,\\
	\langle \mathcal{P}_1^{B_c^*}(^3S_1)\rangle &\simeq & 3\langle \mathcal{P}_1^{B_c}(^1S_0)\rangle=-0.990\,\text{GeV}^5\,.
\end{eqnarray}

We set the FFs at initial renormalization scale $\mu=m_b+m_c=6.00$ GeV as what we present at the end of Sect.\ref{sect:compt}. The corresponding plots are depicted in Fig.\ref{fig:FFsat6GeV}. One can see that the relativistic corrections enhance the distribution around $15\% - 20\%$. 

Since the behavior of the FFs for $b$ to $B_c$, unpolarized $B_c^*$ and transversely polarized $B_c^*$ are similar, for illustration, we just show the effects of the DGLAP evolution for FFs of $B_c$. 
We show the non-singlet DGLAP evolution in Fig.\ref{fig:FFBcNSDGLAP}. One can see that the evolution effects are sizable, but does not change the behavior of $D_{b\to B_c}$ very much. The jump around $z=0$ region is mainly due to the instability of the numerical method. With improving the accuracy of the numerical method, such deficiency would be demmished.  

The singlet DGLAP evolution of $D_{b\to B_c}$, $D_{\bar c \to B_c}$ and $D_{g\to B_c}$ are shown in Fig. \ref{fig:FFbtoBcSDGLAP}, Fig. \ref{fig:FFcbartoBcSDGLAP} and Fig. \ref{fig:FFgtoBcSDGLAP}, respectively. $D_{g\to B_c}(z)$ starts from ${\cal O}(\alpha_s^3)$ which means that $D_{g\to B_c}(z)=0$ at ${\cal O}(\alpha_s^2)$, therefore, for the consistency of the initial condition setting for the RG running, 
here we take $D_{g\to B_c}(z)=0$ at $\mu=6$ GeV as the initial value. From these three diagrams, one can observe that
\begin{itemize}
\item  all three FFs at $z=0$ tend to diverge due to the singlet DGLAP evolution;
\item  the value of $D_{\bar c\to B_c}$ is almost one order of magnitude smaller than $D_{b \to B_c}$, which is understandable since it is harder to get a heavier quark from parton fragmentation;
\item at $\mu=50.0$ GeV and $125.0$ GeV, the values of $D_{g\to B_c}$ are comparable with $D_{\bar c\to B_c}$, which implies that the contribution from gluon fragmentation is needed for more accurate theoretical prediction on $B_c$ production at large transverse momentum region.

\end{itemize}

\section{Summary and outlook\label{sect:summary}}
In this paper, we calculate relativistic corrections to the FFs for a heavy quark to $B_c$ and unpolarized and transversely polarized $B_c^*$ within the framework of NRQCD factorization at the leading order of $\alpha_s$. We show that by adopting the Gremm-Kapustin relation, we can reproduce the results in the literature, however we argue that our scheme to count the relativistic corrections is more consistent in the phenomenological applications. 

We show the ${\cal O}(v^2)$ corrections is around $15-20\%$ as we expect, which is as important as the NLO radiative corrections in \cite{Zheng:2019gnb}. We also show the DGLAP evolution is equally important for phenomenological application. The singlet DGLAP evolution can drastically change the behavior of the FFs of a heavy quark to $B_c^{(*)}$ at the small $z$ region. Furthermore, the singlet DGLAP evolution implies that the FFs of gluon to $B_c^{(*)}$ could be as important as both relativistic corrections and NLO radiative corrections to the FFs of heavy-quark to $B_c^{(*)}$, especially in the production at hadron collider at the large transverse momentum region. Therefore, this FF deserves the further investigation beyond the level of DGLAP evolution. 

When the transverse momentum goes down, the so-called double parton fragmentation mechanism becomes more and more important to increase the theoretical predictive power \cite{Ma:2013yla}. The color-singlet double parton fragmentation function is closely related to the light-cone distribution amplitude (LCDA) which appears mostly in the exclusive production. The LCDAs of $B_c$ meson have been studied up to NLO in both $\alpha_s$ and $v$-expansion in  \cite{Xu:2016dgp,Wang:2017bgv}. However, the color-octet double fragmentation functions for $B_c^{(*)}$ are still missing, and valuable to be carried out.

\section*{Acknowledgement}
We thank Yu Jia and Wen-Long Sang for valuable discussions. 
The work of Deshan Yang is supported in part by the National Natural Science Foundation of China under Grants No.~11275263 and 11635009.

\end{document}